\documentclass[a4paper,twoside,11pt]{article}
\usepackage{a4}
\usepackage{graphics}

\begin{document}
\textheight23cm
\textwidth15.5cm
\oddsidemargin0.5cm
\evensidemargin0cm
\parskip1ex
\pagestyle{myheadings}

\def\sun{\hbox{$\odot$}}
\def\la{\mathrel{\mathchoice {\vcenter{\offinterlineskip\halign{\hfil
$\displaystyle##$\hfil\cr<\cr\sim\cr}}}
{\vcenter{\offinterlineskip\halign{\hfil$\textstyle##$\hfil\cr
<\cr\sim\cr}}}
{\vcenter{\offinterlineskip\halign{\hfil$\scriptstyle##$\hfil\cr
<\cr\sim\cr}}}
{\vcenter{\offinterlineskip\halign{\hfil$\scriptscriptstyle##$\hfil\cr
<\cr\sim\cr}}}}}
\def\ga{\mathrel{\mathchoice {\vcenter{\offinterlineskip\halign{\hfil
$\displaystyle##$\hfil\cr>\cr\sim\cr}}}
{\vcenter{\offinterlineskip\halign{\hfil$\textstyle##$\hfil\cr
>\cr\sim\cr}}}
{\vcenter{\offinterlineskip\halign{\hfil$\scriptstyle##$\hfil\cr
>\cr\sim\cr}}}
{\vcenter{\offinterlineskip\halign{\hfil$\scriptscriptstyle##$\hfil\cr
>\cr\sim\cr}}}}}
\def\degr{\hbox{$^\circ$}}
\def\arcmin{\hbox{$^\prime$}}
\def\arcsec{\hbox{$^{\prime\prime}$}}
\def\utw{\smash{\rlap{\lower5pt\hbox{$\sim$}}}}
\def\udtw{\smash{\rlap{\lower6pt\hbox{$\approx$}}}}
\def\fd{\hbox{$.\!\!^{\rm d}$}}
\def\fh{\hbox{$.\!\!^{\rm h}$}}
\def\fm{\hbox{$.\!\!^{\rm m}$}}
\def\fs{\hbox{$.\!\!^{\rm s}$}}
\def\fdg{\hbox{$.\!\!^\circ$}}
\def\farcm{\hbox{$.\mkern-4mu^\prime$}}
\def\farcs{\hbox{$.\!\!^{\prime\prime}$}}
\def\fp{\hbox{$.\!\!^{\scriptscriptstyle\rm p}$}}

\newcommand{\D}{\displaystyle} 
\newcommand{\T}{\textstyle} 
\newcommand{\SC}{\scriptstyle} 
\newcommand{\SSC}{\scriptscriptstyle} 

\def\AJ{{\it Astron. J.} }
\def\ARAA{{\it Annual Rev. of Astron. \& Astrophys.} }
\def\ApJ{{\it Astrophys. J.} }
\def\ApJL{{\it Astrophys. J. Letters} }
\def\ApJS{{\it Astrophys. J. Suppl.} }
\def\ApP{{\it Astropart. Phys.} }
\def\AA{{\it Astron. \& Astroph.} }
\def\AAR{{\it Astron. \& Astroph. Rev.} }
\def\AAL{{\it Astron. \& Astroph. Letters} }
\def\JGR{{\it Journ. of Geophys. Res.}}
\def\JHEP{{\it Journal of High Energy Physics} }
\def\JPhG{{\it Journ. of Physics} {\bf G} }
\def\PhFl{{\it Phys. of Fluids} }
\def\PR{{\it Phys. Rev.} }
\def\PRD{{\it Phys. Rev.} {\bf D} }
\def\PRL{{\it Phys. Rev. Letters} }
\def\Nature{{\it Nature} }
\def\MNRAS{{\it Month. Not. Roy. Astr. Soc.} }
\def\ZA{{\it Zeitschr. f{\"u}r Astrophys.} }
\def\ZFN{{\it Zeitschr. f{\"u}r Naturforsch.} }
\def\etal{{\it et al.}}

\hyphenation{mono-chro-matic  sour-ces  Wein-berg
chang-es Strah-lung dis-tri-bu-tion com-po-si-tion elec-tro-mag-ne-tic
ex-tra-galactic ap-prox-i-ma-tion nu-cle-o-syn-the-sis re-spec-tive-ly
su-per-nova su-per-novae su-per-nova-shocks con-vec-tive down-wards
es-ti-ma-ted frag-ments grav-i-ta-tion-al-ly el-e-ments me-di-um 
ob-ser-va-tions tur-bul-ence sec-ond-ary in-ter-action
in-ter-stellar spall-ation ar-gu-ment de-pen-dence sig-nif-i-cant-ly
in-flu-enc-ed par-ti-cle sim-plic-i-ty nu-cle-ar smash-es iso-topes 
in-ject-ed in-di-vid-u-al nor-mal-iza-tion lon-ger con-stant
sta-tion-ary sta-tion-ar-i-ty spec-trum pro-por-tion-al cos-mic
re-turn ob-ser-va-tion-al es-ti-mate switch-over grav-i-ta-tion-al
super-galactic com-po-nent com-po-nents prob-a-bly cos-mo-log-ical-ly
Kron-berg}
\def\simle{\lower 2pt \hbox {$\buildrel < \over {\scriptstyle \sim }$}}
\def\simge{\lower 2pt \hbox {$\buildrel > \over {\scriptstyle \sim }$}}


\begin{center}
{{ORIGIN OF COSMIC MAGNETIC FIELDS}}\\
\vskip1.0cm
Peter L. Biermann$^{1,2}$ and Cristina F. Galea$^3$\\
\vskip1.0cm
$^1$Max-Planck Institute for Radioastronomy\\
and\\
$^2$Department of Physics and Astronomy,\\
University of Bonn, Bonn, Germany\\
$^3$Experimental High Energy Physics Department, University of Nijmegen,\\
Nijmegen,The Netherlands \\ \vskip1.0cm
www.mpifr-bonn.mpg.de/div/theory\\
plbiermann@mpifr-bonn.mpg.de, cristina@hef.kun.nl
\end{center}

%
     
\section{Abstract}

We propose that the overlapping shock fronts from young supernova
remnants produce a locally unsteady, but globally steady large scale
spiral shock front in spiral galaxies, where star formation and 
therefore
massive star explosions correlate geometrically with spiral structure. 
This global shock front with its steep gradients in temperature,
pressure and associated electric fields will produce drifts, which in
turn give rise to a strong sheet-like electric current, we propose.  
This
sheet current then produces a large scale magnetic field, which is
regular, and connected to the overall spiral structure.  This 
rejuvenates
the overall magnetic field continuously, and also allows to understand
that there is a regular field at all in disk galaxies.  This proposal
connects the existence of magnetic fields to accretion in disks.  We 
not
yet address all the symmetries of the magnetic field here; the picture
proposed here is not complete.  X-ray observations may be able to test 
it
already.

\section{Introduction}

Magnetic fields are found almost everywhere in the Universe, and 
usually
they are strong, but rarely so strong as to dominate locally, as, e.g., 
in the environment of pulsars.  After magnetic fields had been 
predicted
to exist in the disk of our Galaxy inside the interstellar medium on 
the
basis of cosmic ray arguments, \cite{LBS50,LBS51,Kronberg00}, and the
presence of optical polarization from dust scattering, 
\cite{Spitzer68},
with a strength predicted to be around 5 microGauss, to within 20
percent, they were indeed detected, with a strength as expected, see,
e.g., \cite{Beck96}.  

Now the best measurement for the Solar neighborhood is 6 - 7 
microGauss,
with about 1/3 to 1/2 regular, and 2/3 to 1/2 irregular components,
\cite{Beck96}.

\subsection{Critical observations}

The magnetic field is a special challenge to understand in physics
terms.  Other critical measurements are:

The order of the magnetic field in the Galaxy is destroyed by the
interstellar medium stirring on time scales of order 30 million years, 
as
can be deduced from Cosmic Ray transport arguments.  This is less than
the rotation period of the Galaxy, and so any diffusive process taking
many rotation periods is ruled out.  In contrast, inside stars there is
enough time.

Starburst galaxies such as M82 have a magnetic field, which is ordered
and also in near-equipartition.  And yet the starburst is just a few 
tens
of millions of years old.  The conclusion is once more that the
resurrection time scale of the magnetic field is of order 30 million
years.

It appears that the magnetic field has an overall orientation, with its
direction along the spiral arms inwards always, \cite{Krause98}.  This
seems to be a general pattern for spiral galaxies, as long as they are
not interacting.  There is also a different point of view as regards
our Galaxy, \cite{Han94,Han97,Han99}, but a worry is that local effects
overshadow the global picture for us; and so many expect that also in 
our
Galaxy the magnetic field is pointed globally along the spiral arms
inwards.  However, we have to keep in mind that the observational
statistics are still small.

Galaxies which have a strong magnetic field, but also solid body 
rotation
over that radial range of the disk where the magnetic field has been
measured, are another important test.

Clusters of galaxies have a magnetic field which normally appears to be
quite strong, with a lower limit of several percent of equipartition in
energy density, \cite{CKB99,CKB01}.  There are two examples of
clusters which seem to have fields close to equipartition, e.g.
\cite{EBKW97}.  Even large scale sheets in the distribution of galaxies
have been successfuly measured to have magnetic fields, \cite{Kim89}; 
in
that case the strength is not very certain, between 10 and 100 
nanoGauss
range the best estimates, but the fields could also be stronger; just 
the
existence is certain, since that argument is based on the observed
synchrotron emission.

\subsection{History and reviews}

The history of this topic is large:  The dynamo process was invented
twice, by Steenbeck and his associates, in
\cite{StKR1,StKR2,StKR3,StKR4,StKR5,StKR6,StKR7,StKR8,StKR9},
\cite{StKR10,StKR11,StKR12,StKR13,StKR14}
and by Parker, in
\cite{ENP1,ENP2,ENP3,ENP4,ENP5,ENP6,ENP7,ENP8,ENP9,ENP10,ENP11,ENP12}
and \cite{ENP13,ENP14,ENP15}.  Recent reviews are in
\cite{Krause93,Kronberg94,Beck96,Kulsrud99,Kronberg00,Kronberg01}.  
Large
scale magnetic fields are discussed in
\cite{Vallee90,Kulsrud97,RKB98,Blasi98,Kronberg99,Blasi99a,Birk00,Voelk00}.

\subsection{The challenge}

On the basis of the data the challenge is to find a mechanism to 
generate
an ordered magnetic field which is fast.  We outline a proposal here 
that
may be able to do this.  First attempts along these lines were made in
\cite{EriceB01,Galea02}.  This work reported here is another step along
the way, but we are nowhere near finished.

\section{The equations for mass and current}

We follow the arguments of \cite{LBS50,LBS51,Cowl53,Spitzer62,MR62}, in
the notation of \cite{RL79}.   

Ions have mass $m_i$, charge $Z$, and velocity $v_i$ and electrons mass
$m_e$ and velocity $v_e$, ${\bf{E}}$ and ${\bf{B}}$ are the electric 
and
magnetic field.  Ions have a density $n_i$ of with an average charge of
$Z$, and electrons have a density $n_e$.  Then the two equations of
motion for ions and electrons are

\begin{eqnarray}
n_i m_i( \frac{\partial}{\partial t} {\bf{v_i}} + ({\bf{v_i}} \cdot
\nabla) {\bf{v_i}}) = & \cr + n_i m_i \nu_i \Delta {\bf{v_i}} +
n_i m_i (\nu_i/3 + \zeta_i) {\rm grad}\, {\rm div} \, {\bf{v_i}} +
&\cr - 2 n_i m_i {\bf \omega_P} \times {\bf v_i} + n_i Z e ({\bf{E}} +
\frac{{\bf{v_i}}}{c} \times {\bf{B}}) & \cr - \nabla P_i  - n_i m_i
\nabla \Phi + {\bf{P_{ie}}}
\end{eqnarray}

and

\begin{eqnarray}
n_e m_e( \frac{\partial}{\partial t} {\bf{v_e}} + ({\bf{v_e}} \cdot
\nabla) {\bf{v_e}}) = \frac{\sigma_T {\bf{F}} n_e}{c} + & \cr + 
n_e m_e \nu_e \Delta {\bf{v_e}} + n_e m_e (\nu_e/3 + \zeta_e) 
{\rm grad}\, {\rm div} \, {\bf{v_e}} + &\cr  
- 2 n_e m_e {\bf \omega_P} \times {\bf v_e} 
- n_e e ({\bf{E}} + \frac{{\bf{v_e}}}{c}
\times {\bf{B}}) & \cr - \nabla P_e - n_e m_e \nabla \Phi +
{\bf{P_{ei}}}
\end{eqnarray}

\noindent where ${\bf \omega_P}$ is the pattern velocity of any
preferred rotating system of reference, of interest in the case we
consider a spiral stationary pattern.   We need to point out that these
equations describe the forces and so the changes in the orbit of any 
ion
or electron (positron), but they do not contain the steady flow of
ions/electrons, which are carried around in a circle perhaps; this can
happen in the neighborhood of the intense radiation field of an 
accretion
disk around a black hole:  The radiation field pushes on the skin of
interstellar material close to it, and so a charged layer is formed; 
this
charged layer is carried around in its orbit with the material, and so 
a
current is built up.  This current then produces a specific unique
symmetry, as discussed below.

The stress-tensors ${\bf{\Psi_e}}$ and ${\bf{\Psi_i}}$ are replaced 
with
viscosity terms and the pressure gradient, $\Phi$ is the gravitational
potential, ${\bf{F}}$  the radiative flux, acting only on electrons via
Thomson scattering (cross section  $\sigma_T$), and ${\bf{P_{ei}}} =
-{\bf{P_{ie}}}$ is the frictional exchange of momentum between ions and
electrons, while the term involving the viscosity $\nu_i$ and $\nu_e$
allows for accretion disks. 

We assume, that we can ignore the dependence of the density in the
viscosity terms, and that the viscosity is the same for electrons and 
ions
to first order, since it clearly depends on large scale flows, and at 
the
level at which we are initially interested in, not on the microphysics.  
Therefore we set $\nu_i = \nu_e = \nu$ for the kinematic shearflow
viscosity, and also $\zeta_i = \zeta_e = \zeta$ for the specific volume
viscosity.   These last assumptions are made here only for initial
simplicity.  

We define the density $\rho$, flow velocity ${\bf{v_i}}$, 
$\sigma_{i,e}$
the net charge density, and electric current ${\bf{j}}$ in an obvious
way, 

\begin{eqnarray}
\rho = & n_im_i + n_e m_e \cr 
\rho {\bf{v}} = & n_i m_i {\bf{v_i}} + n_e m_e {\bf{v_e}} \cr
{\bf{j}} = & e ( n_i Z {\bf{v_i}} - n_e {\bf{v_e}}) \cr
\sigma_{i,e} = & n_i Z - n_e \cr
n_i {\bf{v_i}} {m_i} (1 + Z \frac{m_e}{m_i}) = &
\rho {\bf{v}} + \frac{m_e}{e} {\bf{j}} \cr 
n_e {\bf{v_e}} \frac{m_i}{Z} (1 + Z \frac{m_e}{m_i}) = & \rho {\bf{v}} - \frac{m_i}{Z e} {\bf{j}} 
\end{eqnarray}

Cosmic Ray particles can be part of these equations, but due to
the two-stream instability \cite{Spitzer62} they cannot flow faster 
than
the local Alfv{\'e}n speed.  There is a possibility to create
secondaries, electron-positron pairs from charged pion decay, arising
from p-p collisions between cosmic ray particles and the interstellar
medium; another way to consider this to allow for charge exchanges,
recombination and ionization:  Let then
$S_e/n_e$ be the rate of such processes.  $S_i$ and $S_e$ is the rate 
of
density increase of ions and electrons, respectively.  

The Maxwell equations are

\begin{eqnarray}
{\rm{div}} {\bf{\, E}} = & 4 \pi e \, \sigma_{i,e} \cr
{\rm{div}} {\bf{\, B}} = & 0 \cr
{\rm{curl}} {\bf{\, E}} = & - \frac{1}{c} \frac{\partial}{\partial t}
{\bf{B}} \cr
{\rm{curl}} {\bf{\, B}} = & \frac{4 \pi}{c} {\bf{j}} + 
\frac{1}{c} \frac{\partial}{\partial t} {\bf{E}}
\end{eqnarray}

The Maxwell equations allow to relate observed magnetic fields to the
underlying symmetries, and electric currents.  In the following we
ignore the electric field inside a shockwave at first.  We add the two
main equations, to obtain the standard equation of motion, including
magnetic fields and currents, and also subtract these equations to 
obtain
the equation of motion for electric currents.

\subsection{Mass and momentum flow}

We consider the galactic disk, in the observer's inertial frame system,
i.e. keep the rotational flow ${\bf{v_{\phi}}}$ with all spatial
gradients.  The equations allow for a wind from the disk through the 
term
$v_z$, a Galactic wind as an analogy to the Solar wind, driven by
Supernova explosions, individual massive star winds, and HII regions.

\begin{eqnarray}
\rho \frac{\partial}{\partial t} {\bf{v}} =  
\frac{\sigma_T {\bf{F}}}{c} \frac{Z}{m_i} \frac{\rho}{1 + Z m_e/m_i} + 
&
\cr + \rho \nu \Delta {\bf{v}} +
\rho (\nu/3 + \zeta) \,
{\rm grad} \, {\rm div} \,{\bf{v}} \, + &\cr - 2 \rho {\bf \omega_P}
\times {\bf v} + {\bf{T_i}} + \sigma_{i,e} e {\bf E} +
\frac{1}{c} \, {\bf{j}} \times {\bf{B}} - \nabla P -\rho \nabla \Phi
\end{eqnarray}

plus non-linear terms

\begin{eqnarray}
{\bf{T_i}} = -{\bf{v_i}} m_i ({\rm{div}} (n_i {\bf{v_i}}) - S_i) - &\cr
-{\bf{v_e}} m_e ({\rm{div}} (n_e {\bf{v_e}}) - S_e) + &\cr + {\bf{v}}
{\rm{div}} (\rho {\bf{v}}) - {\bf{v}} (m_i S_i + m_e S_e) -  & \cr - 
m_i
n_i ({\bf{v_i}} \cdot \nabla) {\bf{v_i}} - & \cr - m_e n_e ({\bf{v_e}}
\cdot \nabla ) {\bf{v_e}} 
\end{eqnarray}

Momentum exchanges ${\bf{P_{ie}}}
+{\bf{P_{ei}}} =0$ balance out, and the total pressure sum is 
$P = P_i + P_e$.

\subsection{Electric current and momentum flow}

The momentum exchange term is
${\bf{P_{ei}}} = e n_e \eta {\bf{j}}$
with $\eta$ the resistivity.  Subtracting the two main initial 
equations
for ions and electrons gives

\begin{eqnarray}
{\bf{T_e}} = -Z {\bf{v_i}}
({\rm{div}} (n_i {\bf{v_i}}) -S_i)) +&\cr + {\bf{v_e}} ({\rm{div}} (n_e
{\bf{v_e}}) - S_e) + &\cr - Z n_i ({\bf{v_i}} \cdot \nabla) {\bf{v_i}} 
+
n_e ({\bf{v_e}} \cdot \nabla) {\bf{v_e}} 
\end{eqnarray}

\begin{eqnarray}
\frac{1}{e} \, \frac{\partial}{\partial t} \, {\bf{j}} = 
{\bf{T_e}} \, + 
\frac{1}{m_e} \nabla P_e - \frac{Z}{m_i} \nabla P_i \, -
&\cr - 2 \frac{1}{e} {\bf \omega_P} \times {\bf j} - \frac{\sigma_T
{\bf{F}}}{m_e c} \frac{Z}{m_i} 
\frac{\rho}{1 + Z m_e/m_i}\, - &\cr - \sigma_{i,e} {\nabla \Phi} + 
\frac{1}{e} \rho (\nu/3 + \zeta) \, {\rm grad}\, {\rm div} \,
\frac{\bf{j}}{\rho} + 
\frac{1}{e} \rho \nu \Delta \, \frac{\bf{j}}{\rho} \, -
 &\cr - \frac{\sigma_{i,e}}{m_e}(1 -
\frac{Z m_e}{m_i}) e {\bf E} + \frac{e Z \rho}{m_e m_i} \, ( {\bf{\, 
E}}
- \eta {\bf{j}} +
\frac{{\bf{v}}}{c} \times {\bf{B}} + &\cr +
\frac{m_i}{Z
\rho e c} (\frac{Z m_e}{m_i} - 1) \, {\bf{j}} \times {\bf{ B}} )
\end{eqnarray}

and with a properly modified term ${\bf{T_{e,\star}}}$:

\begin{eqnarray}
\frac{m_e m_i}{e^2 Z \rho} \, \frac{\partial}{\partial t} \,
{\bf{j}} = \frac{m_e m_i}{e Z \rho} {\bf{T_{e,\star}}} + {\bf{j}}
\frac{S_e}{n_e} \frac{m_e m_i}{e^2 Z \rho} +  & \cr +
\frac{m_i}{e Z \rho} (\nabla P_e - \frac{Z m_e}{m_i}
\nabla P_i)  -  & \cr - \frac{\sigma_T
{\bf{F}}}{e c} \frac{1}{1 + Z m_e/m_i}\, + & \cr +
\frac{m_e m_i}{e^2 Z} (\nu/3 + \zeta) \, {\rm grad}\, {\rm div} \,
\frac{\bf{j}}{\rho} +
\frac{m_e m_i}{e^2 Z} \nu \Delta \frac{\bf{j}}{\rho} \, -  & \cr
- \frac{m_i}{Z \rho} \, \sigma_{i,e}(1 - \frac{Z m_e}{m_i}) {\bf E} -
\eta {\bf{j}}  \, + {\bf{\, E}} +
\frac{{\bf{v}}}{c} \times {\bf{B}} + & \cr - \frac{m_e m_i}{e Z \rho}
\, \sigma_{i,e} {\nabla \Phi} +
\frac{m_i}{Z \rho e c} (\frac{Z m_e}{m_i} - 1) \, {\bf{j}} \times {\bf{
B}} 
\end{eqnarray}

In deriving these equations we have made various simplifying 
assumptions
in order to keep them tractable, as we will discuss below; we have not
insisted to keep all terms involving the lack of charge neutrality.

\subsection{Number estimates}

{}From the saturation argument of \cite{MR62} as applied to
Galaxy, and its interstellar medium (hot phase, detected by ROSAT,
see \cite{ROSAT97}, ASCA observations \cite{Kaneda97}, RXTE 
measurements,
\cite{Valinia98}) we obtain estimates for the various currents and
scales.  

\begin{itemize}

\item{} The ${\rm{curl}}$-operator scale is $10^{23.2}$ cm, from $2 \pi
r$, where $r$ is the radial distance from us to the center of the 
Galaxy,
about 8 kpc.  This we use to infer from Maxwell's equation the average
electric current in the Galaxy, mentioned below.

\item{}  The z-gradient, the $\nabla$-operator, here has a scale of 2 
kpc
= $10^{21.8}$ cm, from the scale of the hot gas, and the
scaleheight of the magneto-ionic disk is 1.8 kpc \cite{SK80}, basically
the same. 

\item{}  The resistivity according to Spitzer, \cite{Spitzer62}, is
of order $10^{-17}$ sec, but highly uncertain; the uncertainty is many
powers of ten.   This resistivity corresponds to the limit that binary
collisions dominate the exchange of momentum, but if the Larmor 
frequency
were to dominate, then the resistivity could be very much higher,
possibly by more than ten orders of magnitude.

\item{}  The hot gas has density is $3 \, 10^{-3}$ cm$^{-3}$, and its
temperature is $4 \, 10^6$ K \cite{ROSAT97}, with a total gas pressure 
of
$10^{-11.4}$ dyn/cm$^2$.

\item{}  The total magnetic field is approximately $10^{-5.2}$ Gauss, 
and
so the total magnetic field pressure is $10^{-11.8}$ dyn/cm$^2$;  we 
have
therefore  approximately $P_{CR} = P_{B}$,   $P_{CR} + P_{B} = 
10^{-11.5}$
dyn/cm$^2$.  

\item{}  The electric current is $j \, = \, 10^{-18.9} \, {\rm cgs \,
units}$.  For the observed symmetry of the magnetic field there must be 
a
current.  The existence of this current is the key to our approach.

\item{}  These numbers are uncertain, probably to within somewhat less
than a factor of 2 (see E. Berkhuijsen in Beck et al. \cite{Beck96}).  
A 
much higher uncertainty derives from the assumption, that everything is
basically smooth.

\end{itemize}

We make the implicit assumption here, that all relevant time scales are 
of order $3 \, 10^7$ years, or longer; we also assume that the topology
of the medium, permeated by magnetic fields, hot and cold gas, and 
cosmic
rays is basically smooth, with no critical different topologies, such
as shocks.  We basically assume at this stage, that shocks exist, but 
do
not drive any relevant part of the system.

\subsection{Symmetries}

In the following we will use the symmetry with respect to the symmetry
plane $z = 0$ as a first guiding principle:

We introduce the following language:

\begin{itemize}

\item{}  We will call a quantity, vector component or scalar 
``even$(z, z^n)$" in the case, that the quantity can be expanded into a
power series in $z$, that contains only even terms, and starts with 
$z^n$,
where $n$ is an even number.  We will use the expression even$(z,1)$ in
the case that the quantity is a constant in its first term, so $n=0$.

\item{}  We will call a quantity, vector component or scalar 
``odd$(z, z^n)$" in the case, that the quantity can be expanded into a
power series in $z$, that contains only odd terms, and starts with 
$z^n$,
where $n$ is an odd number.

\item{}  We will have to check whether any argument forces us to use a
mixed expresssion - but we have not found one yet.

\item{}  We will use the language (even, even, odd), when the three
vector components in cylindrical coordinates $r, \phi, z$ are in that
sequence even, even, and odd, or, e.g., (odd, odd, even) in a
different case.

\end{itemize}

Looking at Maxwell's equations and the data, we find then that the
main symmetries of different quantities are as follows:

\begin{itemize}

\item{}  ${\bf B}$ is (even, even, odd).

\item{}  ${\bf j}$ is (odd, odd, even).

\item{}  In the observer frame ${\bf E}$ is also (odd, odd, even), as
${\bf j}$.  ${\bf E}$ derives from a Lorentz transformation of a shock
frame ${\bf B}$-field, and so is not an independent source term.

\item{}  The expression ${\bf v} \times {\bf B}$ is (odd, odd, even).

\item{}  The expression $ {\bf j} \times {\bf B}$ is (even, even, odd).

\item{}  The expression $\nabla P$ is (even, even, odd).

\end{itemize}

There is an alternate pattern, in which the magnetic field shows (odd,
odd, even), and the electric current shows (even, even, odd) symmetry. 
It cannot be excluded, that some galaxies have this symmetry.  In the
following we will pursue the discussion of the main symmetry.

What we find then in the case of the equations for mass flow and 
electric
current flow is that the following is true:

\begin{itemize}

\item{}  In the equation for mass flow we have only terms that are 
either
independent of ${\bf j}$, or ${\bf B}$, or quadratic.  All these terms
have the symmetry (even, even, odd).  There are some linear 
cross-terms,
which should cancel each other out; they cannot act as source terms due
to a failure to match symmetries.

\item{}  In the equation for the electric current the entire equation
splits up into terms, that are either (odd, odd, even), or (even, even,
odd).  Those terms which are linear in either ${\bf j}$, ${\bf E}$ or
${\bf B}$ are all (odd, odd, even). Those terms which are either a 
source
term, or quadratic, are all (even, even, odd).  

\item{}  Interestingly, the terms involving the pressure gradient, and
also the Hall term, which are of similar numerical magnitude, do have 
the
same symmetry as well.  They do ``belong together".

\end{itemize}

The first conclusion is then, with the equations for mass flow, and the
electric current as written here, the equations are independent of the
(+,-) symmetry with respect to the sign of ${\bf B}$ and ${\bf j}$. 
Replacing both ${\bf B}$ and also ${\bf j}$ with the same numbers, but
with all signs changed to the opposite, does not change the equation.  
We
conclude, that the physics which determines the sign is not yet
adequately described by these equations.

\subsection{Battery effect and dynamo action}

Now two arguments can be made:

\subsubsection{The dynamo process}

Considering the overall equation for driving a current, and taking the
curl, and dropping all irrelevant terms, one obtains the basic equation
for a discussion of the dynamo process, allowing for Coriolis forces
and turbulence in a rotating system.  The large number of original
references were already given in the introduction.

\subsubsection{Battery effect}

Assuming no magnetic field to exist a priori, one obtains the Battery
effect, \cite{LBS50,LBS51}.  Again taking the ${\rm{curl}}$ of the 
entire
equation for the electric current, we have one important term

\begin{equation}
{\rm{ curl}} \frac{m_i}{e Z \rho} \nabla P_e \sim \nabla (\frac{m_i}{e 
Z
\rho}) \times \nabla P_e 
\end{equation}

In an accretion disk, where density and temperature are usually 
governed
by very different microphysics, this term is certainly non-zero, and
drives an electric current.  This the original effect pointed out by 
the
early authors \cite{LBS50,LBS51}.  This is a slow process, although in 
its
limit it gives the reasonable estimates for magnetic fields, see
\cite{Cowl53}.  The key argument here is that an electric current 
drives
the generation of the magnetic field; this electric current is driven 
by
the rotation of the system.

\subsection{Scaling the hierarchy}

Using the observations, and Maxwell's equations we can discern the
numerical relevance of all the terms in the two main equations, for 
mass
flow and for the electric current; the numbers have been given above
already.  In this we assume at first that there are no discontinuities, 
to
obtain a first guide line of what the observations are telling us.

{Three levels} of numerical strength exist in the system:

The level structure exists already in the equations connecting 
momentum flow and electric current to electron and ion velocities.  
There
are two terms also at a lower level: the {momentum flow of the electric
current} is lower than the real mass momentum flow by $10^{-14.4}$ and
$10^{-17.7}$, respectively.  As a consequence this structuring  also
exists in the fuller equation above, the equation describing the 
driving
of the electric current. 

\begin{itemize}

\item{}  ${\bf{E}}$ and $\frac{\bf{v}}{c} \times
{\bf{B}}$ of order $10^{-8}$ cgs units, using our Galaxy as the prime
example.

\item{}  Pressure gradients and the Hall term are about $10^{-15}$ to
$10^{-18}$ times smaller.  The radiation driving term is about a
factor of 100 weaker again.

\item{}  In ${\bf{T_e}}$ the $({\bf{j}} \cdot \nabla ){\bf{v}}$-terms, 
the time dependent term, the ionization term (assuming an ionization 
time
scale of $3 \, 10^7$ years as for the time dependence), the viscosity
terms are all  again about $10^{-15}$ to $10^{-18}$ times smaller.  The
resistivity term is nominally a factor of about $10^4$ larger, but very
badly understood.

\end{itemize}

We introduce here a way to ``scale the hierarchy", by noting that there
may be a large scale shock wave pattern in the Galaxy:  

In a spiral galaxy there is a spiral pattern of young massive star
formation, and an associated spiral pattern of supernova explosions of
just those stars.  These supernova explosions produce nearly spherical
shockwaves, that propagate outwards, and can reach quite large 
distances
in the tenuous medium.  It is easy to show that these shockwaves can
overlap in a way that there is always a coherent large scale connected 
net
of shock wave surfaces, locally highly time-dependent, but globally
stationary.  Locally these shock waves are very time-dependent, and 
quite
fast, obviously faster then the speed of sound in the hot medium, but 
the
global pattern is slow, with a speed much less than the speed of sound 
in
the hot medium.  So a comparison with the waves on a beach may be
appropiate, very strong waves, as in the Bretagne on the north-western
shore of France, but limited by the beach itself and the tides, running 
up
the beach as breakers with a speed that children match only with
difficulty.

Let us check on the time scales:  In a galaxy like ours we have about 
one
supernova every 100 years, and maybe 1 supernova every $10^3$ years in
the appropiate mass range, those corresponding to zero age main 
sequence
mass above about 15 to 20 solar masses.  This means that a supernova
expands into the tenuous hot interstellar medium to a radius of about 
50
pc within $2.5 \, 10^4$ years, and the shock then still has a velocity 
of
500 km/s, a small multiple of the speed of sound in the hot medium. 
Taking then the notion, that the younger supernovae occur all along the
line of the spiral arms, we have about $10^{-6}$ yr$^{-1}$ and
pc$^{-1}$.  If each supernova reaches out to 100, pc, within $2.5 \,
10^4$ years, then we have at this frequency about 2 supernovae 
occurring
per 100 pc and per this time, so just a sufficient number.

We need to inquire about the ``front" and the ``back" of this
geometrical set of exploding stars:  There is a minimum age for stars 
to
explode as supernovae, but there is a long tail of larger ages, and so 
one can expect that there is a front pattern, but no back-pattern.
At the backside of the spiral pattern there is no corresponding shock;
everything peters out.

This concept is similar to the concept for the young supernova shock
itself, as in \cite{CRI}.

In such a shock an electric field exists, driven by the fact that ions
and electrons have a different mass. Then the overall global spiral 
shock
drives a current system, a sheet current, and the current is very 
strong
inside the sheet, but weak elsewhere.  We note that a sheet current
produces a homogeneous magnetic field, with a different sign on 
opposing
sides.  The current may be driven by the ${\bf E} \, \times \, {\bf B}$
drift inside the shock sheet; also, the gradient in pressure and/or
temperature in this shock can drive a current.  Once one sign is
established, it may prevail due to the overall structure winning by
emphasizing the common vector components, i.e. those geometrically
aligned with the overall shock structure. Since an ${\bf E}
\, \times \, {\bf B}$ drift pushes positive and negative charges in the
same direction, only inertia can produce a real electric current;
inertia is only important if the velocities reach close to thermal
velocities (or relativistic, unlikely here).  The requirement to 
transport
angular momentum transport implies that the sign permeates everything 
in
the same way, since any region, where the sign changes would violate 
the
strict direction of angular momentum transport.

Does this solve the problem?

It goes part of the way towards solving the problem by making all three
levels equivalent, since the second level was small just by the ratio 
of
the Larmor radius of thermal hot ions to the overall scale, and a shock
wave pushes the strength of these terms up by exactly this factor. 
With a strong sheet-like electric current, stronger by about $10^{14}$,
but geometrically limited to the shock sheet, we have then equivalence
between the first and the second level.  And again the third level was 
smaller than the second by again the same factor.  The third level
involves a spatial derivative and the current, and so is doubly pushed 
up
in level.  Therefore, only in the shock sheet and its immediate
environment do we have equivalence of all three levels, previously
disjunct.  With a shock wave all three levels are becoming equivalent
in numerical strength; however, the shock does not change the 
symmetries.

\subsubsection{The current sheet}

This means that we have to consider whether such a current can actually 
be
carried by the sheet:  the thickness of the sheet is about a postshock
thermal Larmor radius $r_g$, and so about $10^9$ cm, stronger than an
average current by the ratio of length scales, $\pi r$ vs. $r_g$.  So 
the
electric current in the sheet is about $10^{-5}$ cgs units.  This means
that the drift velocity is required to be about thermal, so all ions 
and
electrons participate, with a velocity difference about equal to the
thermal velocity.  This thermal velocity here is the preshock velocity,
so the drift can be slow compared to the postshock velocity; however, 
in
order to get a current, we require that electron and ion velocities are
quite different.  We emphasize again, that this very strong electric
current is only in the shock-sheet, and in order to provide the 
magnetic
field observed, has to go vertically.

\subsubsection{Angular momentum transport}

In this model of expanding supernova remnants angular momentum is 
mixed,
and exchanged, which means that it is effectively transported.  The
transport is with the azimuthal velocity difference across a supernova
remnant, and with the diameter of the supernova remnant.  And as shown 
by
Lynden-Bell \& Pringle, \cite{LBP74}, and by Duschl \etal, 
\cite{DSB00},
this leads to an angular momentum transport that scales with some
fraction of the radius $r$, and the azimuthal velocity $v_{\phi}$, with 
a
factor of order 0.01.  This picture developed here does just that.  So
this way of using large scale shock waves, locally highly 
non-stationary
but globally stationary shock waves, does transport angular momentum. 
Whether this is already at the level implied by the observations is not
clear; the numbers suggest that this mechanism is not sufficient.  We
need an additional enhancement.

\subsection{What determines the sign?}

Now we have found that even with all the details we do not get a sign
prevailing in the detailed description of the flow; this is exactly the
same conclusion that Krause \& Beck came to in their original work on 
the
sign of magnetic fields in disk galaxies.

\subsubsection{The role of the central black hole}

Here we wish to explore one way to obtain a specific sign:  In the case 
that a powerfully radiating accretion disk surrounds a central massive
black hole, then we have a strongly ionizing ultraviolet photon field. 
This photon field will ionize and actually charge up a skin in the
surrounding matter.  Carrying this skin of net charge around produces a
circular electric current, with the flow direction exactly the same as
the rotation of the material, presumably with the same sense of 
rotation
as the central black hole.  This will then produce in turn a magnetic
field with a specific sign, aligned with the rotation axis of the black
hole, and pointing in the direction of the spin vector. Related 
questions
have been explored in full relativistic treatment by \cite{KhC94}.

This then may determine a definitive sign overall, because it should 
infect the surrounding matter; a wave of a specific sign will then 
propagate outwards from this environment.  At this point we do not
specify how the sign of this inner magnetic field can be turned into a
sign of the electric current further out.  But once a specific sign is
established, it will sustain itself, within the frame work of the
mechanism outlined above.  

{}From this picture there are some predictions and speculations:
a) Galaxies without a central black hole will not show a specific
sign in their magnetic field configuration. b) Galaxies which did not 
have
an accretion event in a very long time, may lose their memory of the
original sign.  Memory may also be lost, at least for a while, in a
merger event with another galaxy, especially if the merger leads to a
spin-flip of the central black hole. c) In galaxies at high redshift we
may be able to observe the propagation of the wave of a specific sign.

However, this mechanism does not really prefer the observed symmetry 
over
another one, in which the azimuthal magnetic field changes sign in
midplane, so a symmetry of the magnetic field which is (odd, odd,
even).  This latter symmetry is the one that is implicit in any 
symmetry
coming from the central black hole and a charge distribution carried
around in corotation.  The effect of the radiation of the black hole
prefers the alternate symmetry introduced above.

\subsubsection{Transport processes?}

So we have to look at another option, really a program of 
investigation:

Considering how the basic equation was derived from the Boltzmann
equation, we realize that the stress tensor, the heat flux tensor, and
the transport of angular momentum all are dumped into some convenient
functions, and simple dependencies that hide the micro-physics.  It
appears clear from the derivation (see, e.g. Spitzer's book on plasma
physics, \cite{Spitzer62}) that we require a vector, and the angular
momentum flux, that is probably non-linear in ${\bf j}$ and/or ${\bf 
B}$,
and has the symmetry (even, even, odd) in the expression for the mass
flux, while a corresponding vector is required, which has the symmetry
(odd, odd, even) in the equation for the electric current. From the 
idea,
that this is Nature's message, it is clear that the resistivity has
to approach the limit of using the gyrofrequency instead of the 
collision
frequency, and that a non-linear behaviour is required, that leads from
both the main symmetry and the alternate symmetry only to the main
symmetry.  Just geometrically, the vector expression ${\bf B} \times
{{\rm curl} {\bf j}}$ would do just that; this expression leads both 
the
main symmetry and the alternate symmetry into driving an electric
current with the main symmetry - but the relevance for Nature is at 
this
stage mere speculation.  A physical picture for this microphysics has 
to
be left to a later discussion.  What may appear at the end, is that the
requirement that angular momentum flow is outwards drives the sign of 
the
symmetries, and that we have a highly non-linear expression, using 
shock
structures in a disk to scale the hierarchies.  Shocks may appear if 
the
state of the gas is too extreme in its failure to transport angular
momentum, and magnetic fields may correspondingly appear.  It can be
expected that the expressions and coefficients are not independent, 
just
as in the theory of the thermodynamics of irreversible processes.

\subsection{Where magnetic fields may come from}

Therefore we arrive at following speculative picture of the origin
of  magnetic fields:   Stars inject a fairly strong magnetic field,
which is highly chaotic.  This field is injected with extreme chaos, 
but
from the ram pressure condition for the winds with an Alfv{\'e}nic
Machnumber of order 3, \cite{SB97}, and so probably with only a factor
of order 10 below equipartition.  It is actually open to question, 
whether
more is actually required in an inhomogeneous picture of the 
interstellar
medium.  The sheet current in the global shock pattern keeps
restablishing the strength and the symmetries of the overall regular
field, while all the stellar activity produces all the small scale
perturbances.

The sheet current involves a very large scale shock, locally unsteady,
but globally steady.  This is reminiscent of the early discussions in
spiral shock waves, \cite{Lin69,Shu70a,Shu70b,Shu72}.  We are 
reinventing
a large scale shock, highly unsteady locally, with a very high
maximum local shock velocity, but a rather small pattern velocity.

The global shock proposed here is intimately connected to spiral
structure, and thus to the overall flow of angular momentum outwards in
any accretion disk.  One might speculate whether such a spiral shock
could also occur in accretion disks without star formation, and whether
in that case it could again drive a sheet current, and thus provide a
power source for overall magnetic fields.   Could it be that magnetic
field generation is connected to the accretion phenomenon for ionized
disk as soon as they show a spiral shock structure?

One may ask whether such a shock pattern should not be visible?  And
indeed this is an interesting question worthwile to pursue with X-ray
spectroscopy and high spatial resolution data.  X-rays may provide a
crucial test for this speculative proposal.

\section{Conclusions - Future}

Magnetic fields are present in the universe, usually strong and
relevant.  

Where they come from is uncertain, and an attempt outlined here is
still incomplete, but we hope, we will be able to figure out using just
stars, and the physics of the ionized interstellar medium, how magnetic
fields may be generated.

We have outlined a program of investigating the Boltzmann equation to
resolve the connections between heat flow, angular momentum transport,
and driving currents in an accretion disk.  Such an investigation may
finally yield the sign of the prevailing magnetic field, and if so, 
also
elucidate the physics of angular momentum transport and magnetic field
generation in disks at the same time.

What we have proposed here is to suggest that the overlapping 
shockwaves
from supernova remnants produce a locally unsteady, but globally steady
shock structure, which is rather similar in its pattern with the early
attempts to find a global spiral shock; however, through its local
unsteadiness it is really quite different in its physical properties. 
This global shock structure  drives a sheet current through its 
electric
field and strong gradients in pressure and temperature, which in turn
produces by its overall symmetry a global magnetic field, and so
rejuvenates continuously the overall topology of the Galactic magnetic
magnetic field.

Cosmic magnetic fields are key to the transport of charged particles
throughout the Universe, at any energy.

\section{Acknowledgements}

P.L. Biermann would like to acknowledge the hospitality during his
prolonged stay at the University of Paris VII, in the spring 2002,
offered by his colleagues Norma Sanchez and Hector de Vega, as well as
their hospitality at many meetings in Erice, Paris and Palermo.  Work
with PLB is being supported through the AUGER theory and membership 
grant
05 CU1ERA/3 through DESY/BMBF (Germany); further support for the work 
with
PLB comes from the DFG, DAAD, Humboldt Foundation (all Germany), grant
2000/06695-0 from FAPESP (Brasil) through G. Medina-Tanco, a grant
from KOSEF (Korea) through H. Kang and D. Ryu, a grant from ARC
(Australia) through R.J. Protheroe, and European INTAS/ Erasmus/
Sokrates/ Phare grants with partners V. Berezinsky, M.V. Rusu, and V.
Ureche.  The authors would like to thank R. Beck, E. Berkhuijsen, N. Ikhsanov, 
H.
Kang, P.P. Kronberg, H. Lee, G. Medina-Tanco, D. Mitra, and D. Ryu for
many discussions of these questions.

\end{document}